\documentclass[10pt,a4paper,final,twocolumn]{article}

\usepackage{amsthm}
\usepackage[english]{babel}
\usepackage{bmeurer}
\usepackage{color}
\usepackage[colorlinks=false,%
            pdfkeywords={},%
            pdftitle={A Step-indexed Semantic Model of Types for the Call-by-Name Lambda Calculus},%
            pdfauthor={Benedikt Meurer},%
            pdfsubject={},%
            pdfdisplaydoctitle=true]{hyperref}
\usepackage{mathpartir}

\theoremstyle{definition}
\newtheorem{definition}{Definition}
\theoremstyle{plain}
\newtheorem{lemma}[definition]{Lemma}
\newtheorem{theorem}[definition]{Theorem}

\newcommand{\abstr}[2]{\ensuremath{\lambda{#1}.\,{#2}}}
\newcommand{\app}[2]{\ensuremath{{#1}\,{#2}}}
\newcommand{\fix}[1]{\ensuremath{\mathsf{fix}\,{#1}}}
\newcommand{\Nat}{\ensuremath{\mathsf{Nat}}}
\newcommand{\pair}[1]{\ensuremath{\langle{#1}\rangle}}
\newcommand{\floor}[1]{\ensuremath{\left\lfloor{#1}\right\rfloor}}

\begin{document}

\author{%
  Benedikt Meurer\\
  Compilerbau und Softwareanalyse\\
  Universit\"at Siegen\\
  D-57068 Siegen, Germany\\
  \url{meurer@informatik.uni-siegen.de}
}
\date{}
\title{
  A Step-indexed Semantic Model of Types\\
  for the Call-by-Name Lambda Calculus
}
\maketitle

\begin{abstract}
  Step-indexed semantic models of types were proposed as an alternative to purely syntactic
  safety proofs using subject-reduction. Building upon the work by Appel and others, we
  introduce a generalized step-indexed model for the call-by-name lambda calculus. We also
  show how to prove type safety of general recursion in our call-by-name model.
\end{abstract}

\section{Introduction}
\label{sec:Introduction}

Until recently, the most common way to prove type safety was by a purely syntactic proof
technique called subject-reduction, which was adapted from combinatory logic by Wright
and Felleisen \cite{WrightFelleisen94}. One shows that each step of computation preserves
typability (preservation) and that typable states are safe (progress).

This is not the only way though. Type safety can also be proved with respect to a semantic model. The
semantic approach used in this paper avoids formalizing syntactic type expressions. Instead, one defines 
types as sets of semantic values. Using a technique called \emph{step-indexing}, one then relates terms
to these semantic types, and proves that typability implies safety. Instead of formalizing syntactic
typing judgements, one formulates typing lemmata and proves their soundness with respect to the semantic
model.

\paragraph{Related work}
\label{par:Related_work}

Appel \ETAL introduced step-indexed models in the context of foundational proof carrying code \cite{AppelFelty00}.
While they were primarily interested in low-level languages, they also applied their technique to a pure
call-by-value $\lambda$-calculus with recursive types \cite{AppelMcAllester01}. Our work generalizes the framework
by Appel \ETAL to call-by-name by generalizing ground substitutions to terms instead of just values.

Ahmed \ETAL successfully extended the step-indexed models introduced by Appel \ETAL to general references
and impredicative polymorphism \cite{Ahmed04,AhmedAppelVirga02}. Hritcu \ETAL further extended it to object types,
subtyping and bounded quantified types \cite{Hritcu07,HritcuSchwinghammer09}. They also indirectly considered the
call-by-name $\lambda$-calculus using its well-known encoding in the $\varsigma$-calculus \cite{AbadiCardelli96},
including an encoding of the fixed point combinator \cite{Fisher94}.

\paragraph{Outline}
\label{par:Outline}

In section~\ref{sec:The_language_and_its_small_step_semantics} we present the syntax and small step
semantics of the programming language. Section~\ref{sec:Semantic_types} introduces semantic types
and typing lemmata for the simply typed $\lambda$-calculus, which is extended with recursive types in
section~\ref{sec:Recursive_types}, while section~\ref{sec:General_recursion} considers general recursion
in the simply typed $\lambda$-calculus.

\section{The language and its small-step semantics}
\label{sec:The_language_and_its_small_step_semantics}

The language we consider in this paper is the pure $\lambda$-calculus extended with constants,
the simplest functional language that exhibits run-time errors (closed terms that ``go wrong'').
Its syntax is shown in Figure~\ref{fig:Basic_syntax}.
\begin{figure}[htb]
  \centering
  \begin{tabular}{lrcl}
    Variables: & \multicolumn{3}{l}{$x,y,z,\ldots$} \\
    Constants: & $c$ & $::=$ & $\mathsf{0} \mid \mathsf{1} \mid \ldots$ \\
    Terms: & $a,b$ & $::=$ & $c \mid x \mid \abstr{x}{a} \mid \app{a}{b}$
  \end{tabular}
  \caption{Basic syntax}
  \label{fig:Basic_syntax}
\end{figure}
We write $a[x \mapsto b]$ for the (capture avoiding) substitution of $b$ for all unbound occurrences
of $x$ in $a$. A term $v$ is a \emph{value} if it is a constant $c$ or a closed term of the form
$\abstr{x}{a}$.
\begin{figure}[htb]
  \centering
  \begin{mathpar}
    \inferrule{
    }{
      \app{(\abstr{x}{a})}{b} \to a[x \mapsto b]
    }
    \and
    \inferrule{
      a \to a'
    }{
      \app{a}{b} \to \app{a'}{b}
    }
  \end{mathpar}
  \caption{Small-step semantics}
  \label{fig:Small_step_semantics}
\end{figure}

The operational semantics, as shown in small-step style in Figure~\ref{fig:Small_step_semantics},
is entirely conventional \cite{Pierce02}. We write \mbox{$a_0 \to^k a_k$} if there exists a sequence of $k$ 
steps such that \mbox{$a_0 \to a_1 \to \ldots \to a_k$}. We write \mbox{$a \to^* b$} if \mbox{$a \to^k b$}
for some \mbox{$k \ge 0$}. We say that \emph{$a$ is safe for $k$ steps} if for any sequence \mbox{$a \to^j b$}
of \mbox{$j < k$} steps, either $b$ is a value or there is some $b'$ such that \mbox{$b \to b'$}. Note that
any term is safe for $0$ steps. A term $a$ is called \emph{safe} it is safe for every \mbox{$k \ge 0$}.

\section{Semantic types}
\label{sec:Semantic_types}

In this section we construct the methods for proving that a given term is safe in the call-by-name
$\lambda$-calculus, using a simplified type system without recursive types. The semantic approach
taken here considers types as indexed sets of values rather than syntactic type expressions.

\begin{definition} \label{def:Type}
  A \emph{type} is a set $\tau$ of pairs $\pair{k,v}$ where $k \ge 0$ and $v$ is a value, and where the
  set $\tau$ is such that, whenever $\pair{k,v} \in \tau$ and $0 \le j \le k$, then $\pair{j,v} \in \tau$.
  For any term $a$ and type $\tau$ we write $a :_k \tau$ if $a$ is closed, and if, whenever $a \to^j b$ for some
  irreducible term $b$ and $j < k$, then $\pair{k-j,b} \in \tau$.
\end{definition}
Intuitively, $a :_k \tau$ means that the closed term $a$ behaves like an element of $\tau$ for $k$ steps
of computation. That is, $k$ computation steps do not suffice to prove that $a$ does not terminate with
a value of type $\tau$. Note that if $a :_k \tau$ and $0 \le j \le k$ then $a :_j \tau$. Also, for a value
$v$ and $k > 0$, the statements $v :_k \tau$ and $\pair{k,v} \in \tau$ are equivalent.
\begin{definition} \label{def:Typing}
  A \emph{type environment} is a mapping from variables to types. An \emph{environment}
  (or \emph{ground substitution}) is a mapping from variables to terms. For any type environment
  $\Gamma$ and environment $\gamma$ we write \mbox{$\gamma :_k \Gamma$} if \mbox{$\dom(\gamma) = \dom(\Gamma)$} and
  \mbox{$\gamma(x) :_k \Gamma(x)$} for every \mbox{$x \in \dom(\gamma)$}. We write \mbox{$\Gamma \models a :_k \tau$}
  if \mbox{$\gamma(a) :_k \tau$} for every \mbox{$\gamma :_k \Gamma$}, where $\gamma(a)$ is the result of replacing
  the unbound variables in $a$ with their terms under $\gamma$. We write \mbox{$\Gamma \models a : \tau$} if
  \mbox{$\Gamma \models a :_k \tau$} for every \mbox{$k \ge 0$}.
\end{definition}
Note that \mbox{$\Gamma \models a : \tau$} can be viewed as a three place relation that holds on the
type environment $\Gamma$, the term $a$, and the type $\tau$. Utilizing this typing relation
we can express static typing rules, which operate on terms with unbound variables. But first
we observe that the safety theorem, stating ``typability implies safety'', is a direct consequence
of definitions \ref{def:Type} and \ref{def:Typing} now, whereas in a syntactic type theory
it is at least tedious to prove.
\begin{theorem}
  If \mbox{$\emptyset \models a : \tau$}, then $a$ is safe.
\end{theorem}
We can now construct semantic types and appropriate typing lemmata to derive true judgements of the
form \mbox{$\Gamma \models a : \tau$}.
\begin{figure}[htb]
  \centering
  $\begin{array}{rcl}
    \bot &\equiv& \emptyset \\
    \top &\equiv& \{ \pair{k,v} \mid k \ge 0 \} \\
    \Nat &\equiv& \{ \pair{k,c} \mid k \ge 0 \} \\
    \tau \to \tau' &\equiv& \{ \pair{k,\abstr{x}{a}} \mid \forall j < k \forall b.\, \\
    && \quad b :_j \tau \Rightarrow a[x \mapsto b] :_j \tau' \}
    \end{array}$
  \caption{Semantic types}
  \label{fig:Semantic_types}
\end{figure}
Figure~\ref{fig:Semantic_types} gives the types and Figure~\ref{fig:Semantic_typing_lemmata} gives the
typing lemmata for the simply typed $\lambda$-calculus.
The remainder of this section is devoted to proving the soundness of these lemmata.
\begin{figure}[htb]
  \centering
  \begin{mathpar}
    \inferrule{
    }{
      \Gamma \models x : \Gamma(x)
    }
    \and
    \inferrule{
    }{
      \Gamma \models c : \Nat
    }
    \and
    \inferrule{
      \Gamma \models a : \tau \to \tau' \\
      \Gamma \models b : \tau
    }{
      \Gamma \models \app{a}{b} : \tau'
    }
    \and
    \inferrule{
      \Gamma[x \mapsto \tau] \models a : \tau'
    }{
      \Gamma \models \abstr{x}{a} : \tau \to \tau'
    }
  \end{mathpar}
  \caption{Semantic typing lemmata}
  \label{fig:Semantic_typing_lemmata}
\end{figure}

The lemma for variables, stating \mbox{$\Gamma \models x : \Gamma(x)$}, follows directly from the definition of
$\models$. The fact that $\Nat$ is a type, and \mbox{$\Gamma \models c : \Nat$},
both follow immediately from the definition of $\Nat$. We now consider the lemmata for
applications and lambda terms. First we have the following lemma which follows immediately
from the definition of $\to$.
\begin{lemma}
  If $\tau$ and $\tau'$ are types then \mbox{$\tau \to \tau'$} is also a type.
\end{lemma}

\begin{proof}
  By definition of $\to$ it is obvious that \mbox{$\tau \to \tau'$} is closed
  under decreasing index.
\end{proof}

\begin{lemma} \label{lem:Application}
  If \mbox{$a_1 :_k \tau \to \tau'$} and \mbox{$a_2 :_k \tau$}, then \mbox{$(\app{a_1}{a_2}) :_k \tau'$}.
\end{lemma}

\begin{proof}
  Since \mbox{$a_1 :_k \tau \to \tau'$} and \mbox{$a_2 :_k \tau$} we have that both $a_1$ and $a_2$ are closed,
  and if $a_1$ generates an irreducible term in less than $k$ steps, that term must be a lambda
  term. Hence, the application \mbox{$\app{a_1}{a_2}$} either reduces for $k$ steps without any top-level
  $\beta$-reduction, or there must be a lambda term \mbox{$\abstr{x}{b}$} such that
  \mbox{$\app{a_1}{a_2} \to^j \app{(\abstr{x}{b})}{a_2}$} for some \mbox{$j < k$}.

  In the first case, we know that \mbox{$\app{a_1}{a_2}$} is closed, and does not generate an irreducible
  term in less than $k$ steps, and hence \mbox{$\app{a_1}{a_2} :_k \tau'$}.

  Otherwise we have \mbox{$a_2 :_{k-(j+1)} \tau$} by closure under decreasing index, and
  \mbox{$\pair{k-j,\abstr{x}{b}} \in \tau \to \tau'$} by Definition~\ref{def:Type}.
  \mbox{$b[x \mapsto a_2] :_{k-(j+1)} \tau'$} follows by definition of $\to$. But now we have
  \mbox{$\app{a_1}{a_2} \to^{j+1} b[x \mapsto a_2]$} and \mbox{$b[x \mapsto a_2] :_{k-(j+1)} \tau$}, and
  we can conclude $\app{a_1}{a_2} :_k \tau'$.
\end{proof}

\begin{theorem}[Application] \label{thm:Application}
  Let $\Gamma$ be a type environment, let $a_1$ and $a_2$ be (possibly open) terms, and let
  $\tau$ and $\tau'$ be types. If \mbox{$\Gamma \models a_1 : \tau \to \tau'$} and
  \mbox{$\Gamma \models a_2 : \tau$}, then \mbox{$\Gamma \models \app{a_1}{a_2} : \tau'$}.
\end{theorem}

\begin{proof}
  By Lemma~\ref{lem:Application} we have \mbox{$\gamma(\app{a_1}{a_2}) :_k \tau'$} for every $k \ge 0$
  and $\gamma$, whenever \mbox{$\gamma :_k \Gamma$},
  \mbox{$\gamma(a_1) :_k \tau \to \tau'$} and \mbox{$\gamma(a_2) :_k \tau$}. Hence, we conclude
  \mbox{$\Gamma \models \app{a_1}{a_2} :_k \tau'$} (for every $k \ge 0$).
\end{proof}

\begin{theorem}[Abstraction] \label{thm:Abstraction}
  Let $\Gamma$ be a type environment, let $\tau$ and $\tau'$ be types, and let $\Gamma[x \mapsto \tau]$
  be the type environment that is identical to $\Gamma$ except that it maps $x$ to $\tau$. If
  \mbox{$\Gamma[x \mapsto \tau] \models a : \tau'$}, then
  \mbox{$\Gamma \models \abstr{x}{a} : \tau \to \tau'$}.
\end{theorem}

\begin{proof}
  Let $k \ge 0$, $b$ be a closed term with $b :_k \tau$, and $\gamma$ be an environment such
  that \mbox{$\gamma :_k \Gamma$}. Then \mbox{$\gamma[x \mapsto b] :_k \Gamma[x \mapsto \tau]$}, and since
  \mbox{$(\gamma[x \mapsto b])(a) :_k \tau'$} and $b$ is closed, we also have \mbox{$\gamma(a[x \mapsto b]) :_j \tau'$}
  and $b :_j \tau$ for every $j < k$.
  Then \mbox{$\pair{k,\gamma(\abstr{x}{a})} \in \tau \to \tau'$},
  and since \mbox{$\gamma(\abstr{x}{a})$} is obviously closed, we conclude
  \mbox{$\Gamma \models \abstr{x}{a} :_k \tau \to \tau'$} (for every $k \ge 0$).
\end{proof}

\section{Recursive types}
\label{sec:Recursive_types}

Recursive types were one of the main motivations behind the model of Appel \ETAL \cite{AppelMcAllester01}, and
their results apply here, therefore we do not go into much detail. Figure~\ref{fig:Recursive_types}
shows the recursion type operator $\mu$, which computes a candidate fixed point of a function $F$
from types to types by repeatedly applying the function to $\bot$, and the two typing lemmata
for recursive types.
\begin{figure}[htb]
  \centering
  \begin{mathpar}
    \begin{array}{rcl}
      \mu F &\equiv& \{ \pair{k,v} \mid \pair{k,v} \in F^{k+1}(\bot) \} \\
    \end{array}
    \\
    \inferrule{
      \Gamma \models a : F(\mu F)
    }{
      \Gamma \models a : \mu F
    }
    \and
    \inferrule{
      \Gamma \models a : \mu F
    }{
      \Gamma \models a : F(\mu F)
    }
  \end{mathpar}
  \caption{Recursive types}
  \label{fig:Recursive_types}
\end{figure}

We will show that the typing lemmata in Figure~\ref{fig:Recursive_types} hold in the case where
$F$ is \emph{well founded}. This is achieved by proving $\mu F = F(\mu F)$ for every
well founded $F$, essentially proving that our recursive types are actually equi-recursive types,
in contrast to iso-recursive types where $\mu F$ is only isomorphic to $F(\mu F)$ via
\textsf{roll} and \textsf{unroll} constructs on terms \cite{AbadiFiore96,Crary99}.

\begin{definition} \label{def:Approximation}
  The \emph{$k$-approximation} of an indexed set $\tau$ is the subset
  \[ \floor{\tau}_k = \{ \pair{j,v} \mid j < k \wedge \pair{j,v} \in \tau \} \]
  of its elements whose index is less than $k$.
\end{definition}

Obviously $\floor{\tau}_k$ is a type whenever $\tau$ is a type. We now define a notion of well founded
functional. Intuitively, a recursive definition of a type $\tau$ is well founded if, in order to determine
whether or not \mbox{$a :_k \tau$}, it suffices to show \mbox{$b :_j \tau$} for all terms $b$ and indices
\mbox{$j < k$}.

\begin{definition}
  A \emph{well founded functional} is a function $F$ from types to types such that
  \[\floor{F\left(\tau\right)}_{k+1} = \floor{F\left(\floor{\tau}_k\right)}_{k+1}\]
  for every type $\tau$ and every index $k \ge 0$.
\end{definition}

\begin{lemma}
  For every well founded functional $F$ and every $k \ge 0$ we have:
  \begin{enumerate}
  \item $\mu F$ is a type
  \item $\floor{\mu F}_k = \floor{F \left(\mu F\right)}_k$
  \end{enumerate}
\end{lemma}

\begin{theorem} \label{thm:Well_founded_fixpoint}
  If $F$ is a well founded functional, then $\mu F = F(\mu F)$.
\end{theorem}

See the paper of Appel and McAllester \cite{AppelMcAllester01} for the proof sketch.

\section{General recursion}
\label{sec:General_recursion}

As mentioned by Appel \ETAL \cite{AppelMcAllester01}, step-indexed types can also be used to simplify the semantic
treatment of the fixed point rule to type recursive functions in the simply typed $\lambda$-calculus (without
recursive types).
Using the generalized framework presented in section~\ref{sec:Semantic_types}, we are able to provide a direct,
semantic soundness proof of the fixed point rule, which avoids any use of semantic domains, term orders, or
monotonocity.
\begin{figure}[tbh]
  \centering
  \begin{mathpar}
    \inferrule{
    }{
      \fix{a} \to \app{a}{(\fix{a})}
    }
    \and
    \inferrule{
      \Gamma \models a : \tau \to \tau
    }{
      \Gamma \models \fix{a} : \tau
    }
  \end{mathpar}
  \caption{General recursion}
  \label{fig:General_recursion}
\end{figure}

We consider the standard fixed point operator \cite{Pierce02}, written $\fix{a}$, for the
call-by-name lambda calculus. The small step rule and the new typing lemma is shown in
Figure~\ref{fig:General_recursion}. The remainder of this section is devoted to proving the
soundness of the semantic typing lemma.

\begin{lemma}
  If $a :_k \tau \to \tau$, then $(\fix{a}) :_k \tau$.
\end{lemma}

\begin{proof}
  By induction on $k$. Since \mbox{$a :_k \tau \to \tau$} implies \mbox{$a :_j \tau \to \tau$} for every
  \mbox{$j < k$}, we also have \mbox{$(\fix{a}) :_j \tau$} for every \mbox{$j < k$} by induction hypothesis, and using
  Lemma~\ref{lem:Application} we also have \mbox{$(\app{a}{(\fix{a})}) :_j \tau$}. Of course,
  \mbox{$(\fix{a})$} is closed whenever $a$ is closed. So assume that there is some irreducible term
  $b$ and some \mbox{$j < k$} such that \mbox{$(\fix{a}) \to^j b$}.
  This implies \mbox{$j > 0$} and \mbox{$(\fix{a}) \to \app{a}{(\fix{a})} \to^{j-1} b$}, and since
  \mbox{$(\app{a}{(\fix{a})}) :_{j-1} \tau$} we have \mbox{$\pair{k-j,b} \in \tau$}. Hence, we conclude
  \mbox{$(\fix{a}) :_k \tau$}.
\end{proof}

This leads immediately to the following theorem, stating the soundness of the typing lemma
for the call-by-name fixed point operator as shown in Figure~\ref{fig:General_recursion}.

\begin{theorem}[General recursion]
  Let $\Gamma$ be a type environment, let $a$ be a term, and let $\tau$ be a type.
  If \mbox{$\Gamma \models a : \tau \to \tau$}, then \mbox{$\Gamma \models \fix{a} : \tau$}.
\end{theorem}

\section{Conclusion}
\label{sec:Conclusion}

We have presented a step-indexed model for the call-by-name lambda calculus, and used it to
prove the safety of a type system with recursive types. We also proved safety of general
recursion in our framework.

\bibliographystyle{habbrv}
\bibliography{citations}

\end{document}